\documentclass[9pt]{article}
\usepackage[left=10 mm,right=10 mm,top=20 mm,bottom=20 mm]{geometry}
\usepackage{bm}
\usepackage{commath}
\usepackage{xcolor}
\usepackage{soul}
\usepackage{lipsum}
\usepackage{cuted}
\usepackage{etoolbox}
\AfterEndEnvironment{strip}{\leavevmode}
\usepackage{amsmath}
\usepackage{float}
\usepackage[normalem]{ulem}
\usepackage{braket}
\usepackage{amsfonts}
\usepackage{amssymb}
\usepackage{wrapfig}
\usepackage{makeidx}
\usepackage{graphicx}
\usepackage{eqnarray}
\usepackage{caption}
\usepackage{subcaption}
\usepackage{multirow}
\usepackage{tikz}
\usepackage{siunitx}
\usepackage{kbordermatrix}
\usepackage{url}
\usepackage[redeflists]{IEEEtrantools}
\usepackage{array}
\usepackage[nodisplayskipstretch]{setspace}
\usepackage{textcomp}
\usepackage{multicol}

\renewcommand{\kbldelim}{(}
\renewcommand{\kbrdelim}{)}
\allowdisplaybreaks

\providecommand{\keywords}[1]
{
	\small	
	\textbf{\textit{Keywords---}} #1
}

\title{Transition cancellations of $^{87}$Rb and $^{85}$Rb atoms in a magnetic field setting new standards}

\author{Artur Aleksanyan$^{*,1,2}$, Rodolphe Momier$^{2}$, Emil Gazazyan$^{1,3}$, Aram Papoyan$^{1}$, Claude Leroy$^{2}$ \\\\ \small $^1$Institute for Physical Research, NAS of Armenia, Ashtarak-2, 0203, Armenia \\ \small $^2$Laboratoire Interdisciplinaire Carnot de Bourgogne, CNRS UMR 6303, Universit\'e de Bourgogne Franche-Comt\'e, 21000 Dijon, France \\ \small $^3$Yerevan State University, Yerevan, 0025, Armenia \\\\ \small $^*$Corresponding author: arthuraleksan@gmail.com}

\begin{document}
\maketitle

\vspace{15pt}

\begin{abstract}
We have analyzed the magnetic field dependences of intensities of all the optical transitions between magnetic sublevels of hyperfine levels, excited with $\sigma^+$, $\pi$ and $\sigma^-$ polarized light, for the $D_1$ and $D_2$ lines of $^{87}$Rb and $^{85}$Rb atoms. Depending on the type of transition and the quantum numbers of involved levels, the Hamiltonian matrices are of $1\times 1$, $2\times 2$, $3\times 3$ or $4\times 4$ dimension. As an example, analytical expressions are presented for the case of $2\times 2$ dimension matrices for $D_1$ line of both isotopes. Eigenvalues and eigenkets are given, and the expression for the transition intensity as a function of $B$ has been determined. It is found that some $\pi$ transitions of $^{87}$Rb and $^{85}$Rb get completely canceled for certain, extremely precise, values of $B$. No cancellation occurs for $\sigma^+$ or $\sigma^-$ transitions of $D_1$ line. For matrices with size over $2\times 2$, analytical formulas are heavy, and we have performed numerical calculations. All the $B$ values cancelling $\sigma^+$, $\pi$ and $\sigma^-$ transitions of $D_1$ and $D_2$ lines of $^{87}$Rb and $^{85}$Rb are calculated, with an accuracy limited by the precision of the involved physical quantities. We believe our modeling can serve as a tool for determination of standardized values of magnetic field. The experimental implementation feasibility and its possible outcome are addressed. We believe the experimental realization will allow to increase precision of the physical quantities involved, in particular the upper state atomic levels energy.
\end{abstract}

\vspace{10pt}


\vspace{20pt}

\begin{multicols}{2}
\section{Introduction}
Laser spectroscopy of atomic vapors of alkali metals (Na, K, Rb, Cs) is widely used in atomic physics and numerous emerging applications, including quantum information, optical metrology, laser and sensor technologies, etc. \cite{Aleksanyan2020, Legaie2018, Gazazyan2017}. Interest in such single-electron atomic media is caused by the simplicity of energy levels and the presence of strong optical transitions in the visible and near infrared, for which narrow--linewidth cw lasers are widely available. In recent decades, various magneto-optical processes in vapors of alkali metals have been intensively investigated, which is in particular due to interest in the development of new schemes of optical magnetometry \cite{Wilson,Francis,Budker2007}.

Among these processes is modification of the frequency and intensity of optical transitions between individual magnetic sublevels of the hyperfine structure of atoms in a magnetic field. It is well known that in an external magnetic field $B$, the initially degenerate atomic energy levels are split into magnetic sublevels (Zeeman splitting). The corresponding linear shift of atomic transition frequencies with $B$-field holds till it becomes comparable with the hyperfine splitting. With the further increase of the $B$-field, the transition frequencies strongly deviate from the linear behavior \cite{Tremblay,Papageorgiou1994}. Also, significant changes occur for atomic transition probabilities. Further increase of the $B$-field results in re-establishment of linear frequency dependence and stabilization of the transition probabilities (hyperfine Paschen--Back regime) \cite{Sargsyan2014,Sargsyan2018}.

The experimental observation of the above modifications, especially for relatively weak magnetic fields ($\lesssim$~1000~G), is strongly complicated due to the thermal motion of atoms in vapor: individual transitions between the magnetic sublevels are Doppler--broadened (hundreds of MHz), and they overlap under a wide Doppler profile. This complexity can be overcome by using the methods of sub--Doppler spectroscopy, in particular, using optical nanocells \cite{Sargsyan2016,Hakhumyan2012,Klinger2017-2}. It is important to note that in addition to a significant decrease in the inhomogeneous broadening of transitions, the spectroscopy of nanocells (e.g. derivative selective reflection technique \cite{Sargsyan2016, Sargsyan2017, Klinger2017-2}) also allows one to preserve the linear response of the medium (the magnitude of the atomic signal is directly proportional to the transition probability) \cite{Dutier2003, Papoyan2004}.

In recent years, a number of papers have been published devoted to the study of the behavior of atomic transitions in a wide range of magnetic field spanning from the Zeeman to hyperfine Paschen--Back regime (G to kG scale) \cite{Windholz}. Along with the experiment, theoretical models have been developed that give very good agreement with the measurement results. Among other results, strong transitions that are forbidden by the selection rules in a zero magnetic field (magnetically--induced transitions), as well as significant suppression of the initially allowed transitions were observed exploiting different polarizations of the exciting laser radiation \cite{Sargsyan2017-2, Tonoyan2018}.

In this paper we use our theoretical model to determine polarization configurations and magnetic field values, which outright cancel the transitions between individual magnetic sublevels of rubidium atom (i.e. drive the transition probability to zero). The analysis is done for $D_1$ and $D_2$ lines of $^{85}$Rb and $^{87}$Rb. For $D_1$ line, complete analytical and numerical study is done. Analytical formulas for magnetic field values are obtained, which are shown to be in very good agreement with analytically obtained values. For $D_2$ line, the study is done using numerical methods. All the magnetic field values, which cancel the transitions are obtained. This set of values can become a new standard, and may be used to improve the values of physical constants involved in the model.

We also address the issues related to experimental feasibility of the $B$-field cancellation of transitions, and outline the possible applications, such as optical mapping of magnetic field and $B$-field control of optical information.

\section{Theoretical background and analytical example}
\label{sec:theory}
The characteristic polynomial of a $3 \times 3$ or $4 \times 4$ matrix admit analytical expressions for its roots (based on Cardano and Ferrari's formulas), but they are too heavy to be exhibited in an article. Thus in order to explain clearly the way we will determine the magnetic field values, we begin here-after with a $2 \times 2$ matrix for a $\pi$ transition in the case of $^{87}$Rb $D_1$ line.

More precisely we will consider the $\pi$ transitions from $5^2S_{1/2}$ ($F_g=1,2$) to $5^2P_{1/2}$ ($F_e=1,2$) with $m_g=m_e=-1$.

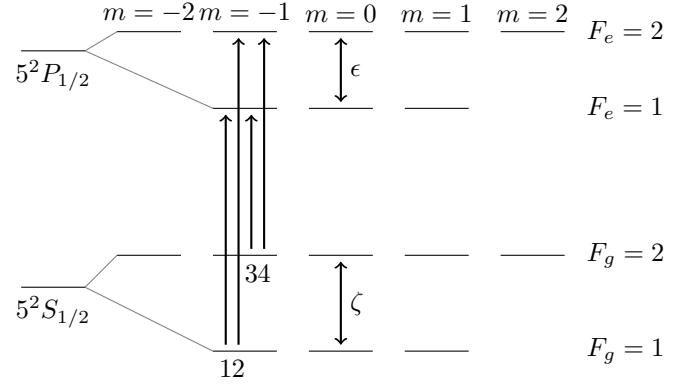
\begin{figure}[H]
\begin{center}
\begin{tikzpicture}[scale=0.85]
\draw[gray] (-0.5,4.7) -- (0,5); \draw[gray] (-0.5,4.7) -- (1.5,3.8); \draw[gray] (-0.5,1) -- (0,1.5); \draw[gray] (-0.5,1) -- (1.5,0);

\draw (0,5) -- (1,5); \node [above] at (0.5,5) {$m=-2$};
\draw (1.5,5) -- (2.5,5); \node [above] at (2,5) {$m=-1$};
\draw (3,5) -- (4,5); \node [above] at (3.5,5) {$m=0$};
\draw (4.5,5) -- (5.5,5); \node [above] at (5,5) {$m=1$};
\draw (6,5) -- (7,5); \node [above] at (6.5,5) {$m=2$};
\node [right] at (7.2,5) {$F_e=2$};

\draw (1.5,3.8) -- (2.5,3.8) (3,3.8) -- (4,3.8) (4.5,3.8) -- (5.5,3.8); \node [right] at (7.2,3.8) {$F_e=1$}; 

\draw (0,1.5) -- (1,1.5) (1.5,1.5) -- (2.5,1.5) (3,1.5) -- (4,1.5) (4.5,1.5) -- (5.5,1.5) (6,1.5) -- (7,1.5); \node [right] at (7.2,1.5) {$F_g=2$}; 
\draw (1.5,0) -- (2.5,0) (3,0) -- (4,0) (4.5,0) -- (5.5,0); \node [right] at (7.2,0) {$F_g=1$}; 

\draw [thick] [->] (1.7,0.1) -- (1.7,3.7); \node [below] at (1.7,0) {1};
\draw [thick] [->] (1.9,0.1) -- (1.9,4.9); \node [below] at (1.9,0) {2};
\draw [thick] [->] (2.1,1.6) -- (2.1,3.7); \node [below] at (2.1,1.5) {3};
\draw [thick] [->] (2.3,1.6) -- (2.3,4.9); \node [below] at (2.3,1.5) {4};

\draw [thick] [<->] (3.5,0.1) -- (3.5,1.4); \node [right] at (3.5,0.75) {$\zeta$};
\draw [thick] [<->] (3.5,3.9) -- (3.5,4.9); \node [right] at (3.5,4.4) {$\epsilon$};

\draw (-1.5,4.7) -- (-0.5,4.7); \node [below] at (-1,4.7) {$5^2P_{1/2}$};
\draw (-1.5,1) -- (-0.5,1); \node [below] at (-1,1) {$5^2S_{1/2}$};
\end{tikzpicture}
\caption[]{$^{87}$Rb $D_1$ line scheme in magnetic field with $\pi$ transitions for $m=-1$.}
\label{fig:87_D1_pi_scheme}
\end{center}
\end{figure}

As shown on Fig.~\ref{fig:87_D1_pi_scheme}, we denote $\zeta$ the frequency difference between the ground state $F_g=2$ and $F_g=1$ levels, and $\epsilon$ the frequency difference between the $F_e=2$ and $F_e=1$ excited state levels.

Accordingly to \cite{Tremblay}, that is with the same quantization axis, in the unperturbed basis $\ket{F,m}$, the diagonal elements of the Hamiltonian matrix $H$ have the following form:
\begin{equation}
\bra{F,m}H\ket{F,m}= E_{0}(F)-\mu_B g_{F} m B,
\label{eq:tremblay1}
\end{equation}
where $E_0(F)$ is the energy of the hyperfine $F$ level, $\mu_B$ is the Bohr magneton, $g_F$ is the associated Land\'e factor, $m$ is the magnetic quantum number and $B$ is the magnetic field ($\vec{B}$ projection on the quantization axis). Non-diagonal elements are given by
\begin{multline}
\bra{F-1,m}H\ket{F,m}=\bra{F,m}H\ket{F-1,m} =-\frac{\mu_B}{2}\left(g_{J}-g_{I}\right) \\
 \times B \left(\frac{\left[(J+I+1)^{2}-F^{2}\right]\left[F^{2}-(J-I)^{2}\right]}{F}\right)^{1/2} \\ 
 \times \left(\frac{F^{2}-m^{2}}{F(2 F+1)(2 F-1)}\right)^{1/2},
\end{multline}
where $g_J$ and $g_I$ are Land\'e factors \cite{HansBook}.
One should note that in what follows one will keep the most exact values of the Land\'e factors as we want to obtain exact analytical relations in the present paper.

Using the above formulas, the ground state and excited state Hamiltonian matrices in the presence of magnetic field are:

\end{multicols}

\setstretch{0.9}
\begin{equation}
\begingroup
\setlength\arraycolsep{-0.62pt}
H_{g}=
\kbordermatrix{\mbox{} & \ket{F_g=2,m_g=-2} & \ket{F_g=2,m_g=-1} & \ket{F_g=1,m_g=-1} & \ket{F_g=2,m_g=0} & \ket{F_g=1,m_g=0} & \ket{F_g=2,m_g=1} & \ket{F_g=1,m_g=1} & \ket{F_g=2,m_g=2} \\
& \zeta+2\mu_B B g_{g} & 0 & 0 & 0 & 0 & 0 & 0 & 0 \\
& 0 & \zeta+\mu_B B g_{g} & \sqrt{3}\mu_B B (g_I-g_{g}) & 0 & 0 & 0 & 0 & 0 \\
& 0 & \sqrt{3}\mu_B B (g_I-g_{g}) & \mu_B B(2 g_I-g_{g}) & 0 & 0 & 0 & 0 & 0 \\
& 0 & 0 & 0 & \zeta & 2\mu_B B (g_I-g_{g}) & 0 & 0 & 0 \\
& 0 & 0 & 0 & 2\mu_B B (g_I-g_{g}) & 0 & 0 & 0 & 0 \\
& 0 & 0 & 0 & 0 & 0 & \zeta-\mu_B B g_{g} & \sqrt{3}\mu_B B (g_I-g_{g}) & 0 \\
& 0 & 0 & 0 & 0 & 0 & \sqrt{3}\mu_B B (g_I-g_{g}) & \mu_B B(g_{g}-2 g_I) & 0 \\
& 0 & 0 & 0 & 0 & 0 & 0 & 0 & \zeta-2\mu_B B g_{g} \\
},
\label{eq:ground_Hamiltonian}
\endgroup
\end{equation}
\begin{equation}
\begingroup
\setlength\arraycolsep{-0.62pt}
H_{e}=
\kbordermatrix{\mbox{} & \ket{F_e=2,m_e=-2} & \ket{F_e=2,m_e=-1} & \ket{F_e=1,m_e=-1} & \ket{F_e=2,m_e=0} & \ket{F_e=1,m_e=0} & \ket{F_e=2,m_e=1} & \ket{F_e=1,m_e=1} & \ket{F_e=2,m_e=2} \\
& \epsilon+2\mu_B B g_{e} & 0 & 0 & 0 & 0 & 0 & 0 & 0 \\
& 0 & \epsilon+\mu_B B g_{e} & \sqrt{3}\mu_B B (g_I-g_{e}) & 0 & 0 & 0 & 0 & 0 \\
& 0 & \sqrt{3}\mu_B B (g_I-g_{e}) & \mu_B B(2 g_I-g_{e}) & 0 & 0 & 0 & 0 & 0 \\
& 0 & 0 & 0 & \epsilon & 2\mu_B B (g_I-g_{e}) & 0 & 0 & 0 \\
& 0 & 0 & 0 & 2\mu_B B (g_I-g_{e}) & 0 & 0 & 0 & 0 \\
& 0 & 0 & 0 & 0 & 0 & \epsilon-\mu_B B g_{e} & \sqrt{3}\mu_B B (g_I-g_{e}) & 0 \\
& 0 & 0 & 0 & 0 & 0 & \sqrt{3}\mu_B B (g_I-g_{e}) & \mu_B B(g_{e}-2 g_I) & 0 \\
& 0 & 0 & 0 & 0 & 0 & 0 & 0 & \epsilon-2\mu_B B g_{e} \\
},
\label{eq:excited_Hamiltonian}
\endgroup
\end{equation}

\begin{multicols}{2}

where the following notations are used:
$g_{g}=\dfrac{3 g_I}{4}+\dfrac{g_S}{4}$ and $g_{e}=\dfrac{3 g_I}{4}+\dfrac{g_L}{3}-\dfrac{g_S}{12}$, where $g_I$, $g_L$ and $g_S$ are respectively nuclear, electron orbital and electron spin Land\'e factors \cite{HansBook}.

Obviously, these two matrices are $m$-block diagonal. As we are interested in transitions, i.e. in difference of energies, all the $E_0(F)$ of \eqref{eq:tremblay1} have been put to zero for the matrices of dimension higher than one. Moreover it allows us to extract the two sub-matrices $G$ and $E$ concerning the $\pi$ transitions $\ket{F_g=1,2,m_g=-1} \longrightarrow \ket{F_e=1,2,m_e=-1}$:
\begin{equation}
\begin{aligned}
\begingroup
\setlength\arraycolsep{2pt}
G=
\kbordermatrix{\mbox{} & \ket{F_g=2,m_g=-1} & \ket{F_g=1,m_g=-1} \\
& \zeta+\mu_B B g_{g} & \sqrt{3}\mu_B B (g_I-g_{g}) \\
& \sqrt{3}\mu_B B (g_I-g_{g}) & \mu_B B(2 g_I-g_{g}) \\
}
\endgroup, \\
\begingroup
\setlength\arraycolsep{2pt}
E=
\kbordermatrix{\mbox{} & \ket{F_e=2,m_e=-1} & \ket{F_e=1,m_e=-1} \\
& \epsilon+\mu_B B g_{e} & \sqrt{3}\mu_B B (g_I-g_{e}) \\
& \sqrt{3}\mu_B B (g_I-g_{e}) & \mu_B B(2 g_I-g_{e}) \\
}
\endgroup
\end{aligned}
\label{eq:E_G_matrices}
\end{equation}
Eigenvalues of the $G$ matrix are given by
\begin{multline}
\lambda_{g\pm}(B)=\dfrac{\zeta+2\mu_B B g_{I}}{2}\\ \pm \dfrac{\sqrt{(\zeta+2\mu_B B (g_{g}-g_{I}))^2+12 \mu_B^2 B^2(g_{g}-g_{I})^2}}{2},
\end{multline}
with the corresponding eigenkets expressed in terms of the unperturbed atomic state vectors
\begin{equation}
\ket{\psi(F_{g}, m_{g})}=\sum_{F_{g}^{\prime}} c_{F_{g} F^{\prime}_{g}}\ket{F_{g}^{\prime}, m_{g}}
\end{equation}
are given by 
\begin{multline}
\ket{\psi(F_{g},m_{g})\pm}=\dfrac{1}{\sqrt{1+\kappa_{g\pm}^2}}\ket{F_g=2,m_{g}=-1}\\
+\dfrac{\kappa_{g\pm}}{\sqrt{1+\kappa_{g\pm}^2}}\ket{F_g=1,m_{g}=-1},
\label{eq:eigket_ground}
\end{multline}
where $\kappa_{g\pm}(B)=\dfrac{\lambda_{g\pm}(B)-\zeta-\mu_B B g_{g}}{\sqrt{3} \mu_B B (g_{I}-g_{g})}$.

Similarly for the $E$ matrix we obtain
\begin{equation}
\begin{aligned}
\lambda_{e\pm}(B) &=\dfrac{\epsilon+2\mu_B B g_{I}}{2} \\
&\pm\dfrac{\sqrt{(\epsilon+2\mu_B B (g_{e}-g_{I}))^2+12 \mu_B^2 B^2(g_{e}-g_{I})^2}}{2},
\end{aligned}
\end{equation}
and the corresponding eigenkets expressed in terms of the unperturbed atomic state vectors
\begin{equation}
\left|\psi\left(F_{e}, m_{e}\right)\right\rangle=\sum_{F_{e}^{\prime}} c_{F_{e} F^{\prime}_{e}}\ket{F_{e}^{\prime}, m_{e}}
\end{equation}
are given by
\begin{multline}
\ket{\psi(F_{e},m_{e})\pm}=\dfrac{1}{\sqrt{1+\kappa_{e\pm}^2}}\ket{F_e=2,m_{e}=-1}\\
+\dfrac{\kappa_{e\pm}}{\sqrt{1+\kappa_{e\pm}^2}}\ket{F_e=1,m_{e}=-1},
\label{eq:eigket_excited}
\end{multline}
where $\kappa_{e\pm}(B)=\dfrac{\lambda_{e\pm}(B)-\epsilon-\mu_B B g_{e}}{\sqrt{3} \mu_B B(g_{I}-g_{e})}$.

The electric dipole component $D_q$ \cite{Tremblay} is determined using the following relation:
\begin{multline}
|\bra{e}D_{q}\ket{g}|^{2}=\frac{3 \epsilon_{0} \hbar \Gamma_{e} \lambda_{e g}^{3}}{8 \pi^{2}} a^{2}[\ket{\psi(F_{e}, m_{e})} ; \ket{\psi(F_{g}, m_{g})} ; q],
\end{multline}
where $\epsilon_{0}$ is the vacuum electric permittivity, $\Gamma_{e}$ is the natural decay rate, $\lambda_{e g}$ is the wavelength between ground and excited states, $q=0,\pm 1$ stands respectively for $\pi$, $\sigma^{\pm}$ transitions. The transfer coefficient reads:
\begin{multline}
a[\ket{\psi(F_{e}, m_{e})} ; \ket{\psi(F_{g}, m_{g})} ; q]
\\ =\sum_{F_{e}^{\prime}, F_{g}^{\prime}} c_{F_{e} F_{e}^{\prime}} a(F_{e}^{\prime}, m_{e} ; F_{g}^{\prime}, m_{g} ; q) c_{F_{g} F_{g}^{\prime}},
\label{eq:a_coeff_sum}
\end{multline} 
where $a(F_{e}, m_{e} ; F_{g}, m_{g} ; q)$ are the unperturbed transfer coefficients:
\begin{multline}
a(F_{e}, m_{e} ; F_{g}, m_{g} ; q) \\ =(-1)^{1+I+J_{e}+F_{e}+F_{g}-m_{e}}\sqrt{2 J_{e}+1}\sqrt{2 F_{e}+1}\sqrt{2 F_{g}+1} \\ \times \left(\begin{array}{ccc}
{F_{e}} & {1} & {F_{g}} \\
{-m_{e}} & {q} & {m_{g}}
\end{array}\right)\left\{\begin{array}{ccc}
{F_{e}} & {1} & {F_{g}} \\
{J_{g}} & {I} & {J_{e}}
\end{array}\right\},
\label{eq:a_coeff_3j_6j}
\end{multline}
which depends on 3-j (parenthesis) and 6-j (curly brackets) symbols.

From the ground eigenstates \eqref{eq:eigket_ground} to the excited eigenstates \eqref{eq:eigket_excited}, four $\pi$ transitions are a priori possible. In order to calculate the values of the magnetic field likely to cancel a transition, it is more relevant to consider the change of sign of the quantity $a[\ket{\psi(F_{e}, m_{e})} ; \ket{\psi(F_{g}, m_{g})} ; q]$ rather than it square. Indeed, it is for an extremely precise value of the magnetic field that a transition is canceled, but via a computer code, whatever the step of variation $\Delta B$ of the field $B$, this precise value of the field $B$ verifying $a^{2}[\ket{\psi(F_{e}, m_{e})} ; \ket{\psi(F_{g}, m_{g})} ; q]=0$ will never be reached. This situation will be even more evident in the case of $3 \times 3$ and $4 \times 4$ matrices, since in these cases, we can hardly hope to obtain simple and compact formulas, function of the variables of our model, and giving the value of the field $B$ which cancels a transition. Only numerical values, also extremely precise can be given and the change of sign of the quantity $a[\ket{\psi(F_{e}, m_{e})} ; \ket{\psi(F_{g}, m_{g})} ; q]$ ensures the nullity of its square.

Coming back to our $2 \times 2$ matrices, let's consider the first quantity $a(\ket{\psi(F_e,m_e)-},\ket{\psi(F_g,m_g)-},q=0)$. From relations \eqref{eq:a_coeff_sum} and \eqref{eq:a_coeff_3j_6j} it reads

\begin{multline}
a(\ket{\psi(F_e,m_e)-},\ket{\psi(F_g,m_g)-},q=0)\\=\dfrac{1}{\sqrt{1+\kappa_{e-}^2}} \times a\left(2,-1;2,-1;0\right) \times \dfrac{1}{\sqrt{1+\kappa_{g-}^2}}\\
+\dfrac{1}{\sqrt{1+\kappa_{e-}^2}} \times a\left(2,-1;1,-1;0\right) \times
\dfrac{\kappa_{g-}}{\sqrt{1+\kappa_{g-}^2}}\\
+\dfrac{\kappa_{e-}}{\sqrt{1+\kappa_{e-}^2}} \times a\left(1,-1;2,-1;0\right) \times \dfrac{1}{\sqrt{1+\kappa_{g-}^2}}\\
+\dfrac{\kappa_{e-}}{\sqrt{1+\kappa_{e-}^2}} \times a\left(1,-1;1,-1;0\right) \times
\dfrac{\kappa_{g-}}{\sqrt{1+\kappa_{g-}^2}}\\
=\dfrac{\sqrt{3}(\kappa_{e-}\kappa_{g-}+\sqrt{3}\kappa_{e-}+\sqrt{3}\kappa_{g-}-1)}{6\sqrt{1+\kappa_{e-}^2}\sqrt{1+\kappa_{g-}^2}}.
\end{multline}
Solving $a(\ket{\psi(F_e,m_e)-},\ket{\psi(F_g,m_g)-},q=0)=0$ leads to
\begin{equation}
B_{(-)}^{(-)}=\dfrac{1}{\mu_B} \cdot \dfrac{3\zeta\epsilon}{3 g_I \epsilon -3 g_S \epsilon + 3 g_I \zeta -4 g_L \zeta +g_S \zeta}.
\label{eq:B--}
\end{equation}
For the second transition, the equation 
\begin{multline}
a(\ket{\psi(F_e,m_e)-},\ket{\psi(F_g,m_g)+},q=0)\\=\dfrac{\sqrt{3}(\kappa_{e-}\kappa_{g+}+\sqrt{3}\kappa_{e-}+\sqrt{3}\kappa_{g+}-1)}{6\sqrt{1+\kappa_{e-}^2}\sqrt{1+\kappa_{g+}^2}}=0
\end{multline}
has no solution. The equation corresponding to the third transition 
\begin{multline}
a(\ket{\psi(F_e,m_e)+},\ket{\psi(F_g,m_g)-},q=0)\\=\dfrac{\sqrt{3}(\kappa_{e+}\kappa_{g-}+\sqrt{3}\kappa_{e+}+\sqrt{3}\kappa_{g-}-1)}{6\sqrt{1+\kappa_{e+}^2}\sqrt{1+\kappa_{g-}^2}}=0
\end{multline}
has no solution too. In the last fourth case, equation
\begin{multline}
a(\ket{\psi(F_e,m_e)+},\ket{\psi(F_g,m_g)+},q=0)\\=\dfrac{\sqrt{3}(\kappa_{e+}\kappa_{g+}+\sqrt{3}\kappa_{e+}+\sqrt{3}\kappa_{g+}-1)}{6\sqrt{1+\kappa_{e+}^2}\sqrt{1+\kappa_{g+}^2}}=0
\end{multline}
leads to the same relation as in the first case (see \eqref{eq:B--})
\begin{equation}
B_{(+)}^{(+)}=\dfrac{1}{\mu_B} \cdot \dfrac{3\zeta\epsilon}{3 g_I \epsilon -3 g_S \epsilon + 3 g_I \zeta -4 g_L \zeta +g_S \zeta}.
\label{eq:B++}
\end{equation}
Formulas \ref{eq:B--} and \ref{eq:B++} are analogous to the ones determined by Momier \textit{et al.} \cite{Momier} for the $5^{2}S_{1/2}\rightarrow6^{2}P_{1/2,\;3/2}$ transitions of $^{87}$Rb.
	
\section{Numerical simulation and comparison with analytically obtained values of rubidium $D_1$ line transition cancellations}
\label{sec:D_1_line}
\begin{figure}[H]
\centering
\includegraphics[scale=0.7]{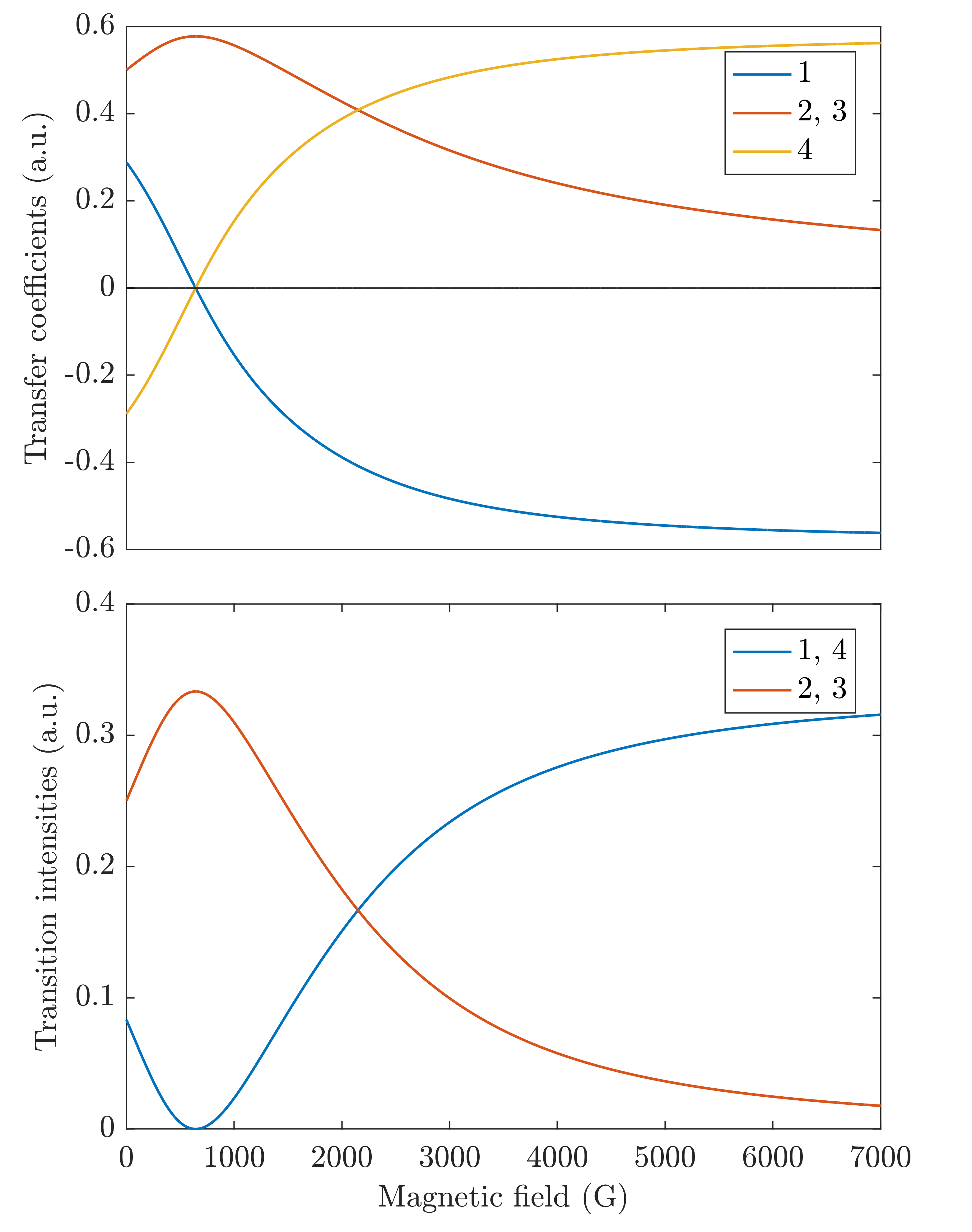}
\caption[]{$^{87}$Rb $D_1$ line $\pi$ transition transfer coefficients and transition intensities for $m=-1$ magnetic quantum number.}
\label{fig:87_D1_pi}
\end{figure}

\begin{figure}[H]
\centering
\includegraphics[scale=0.7]{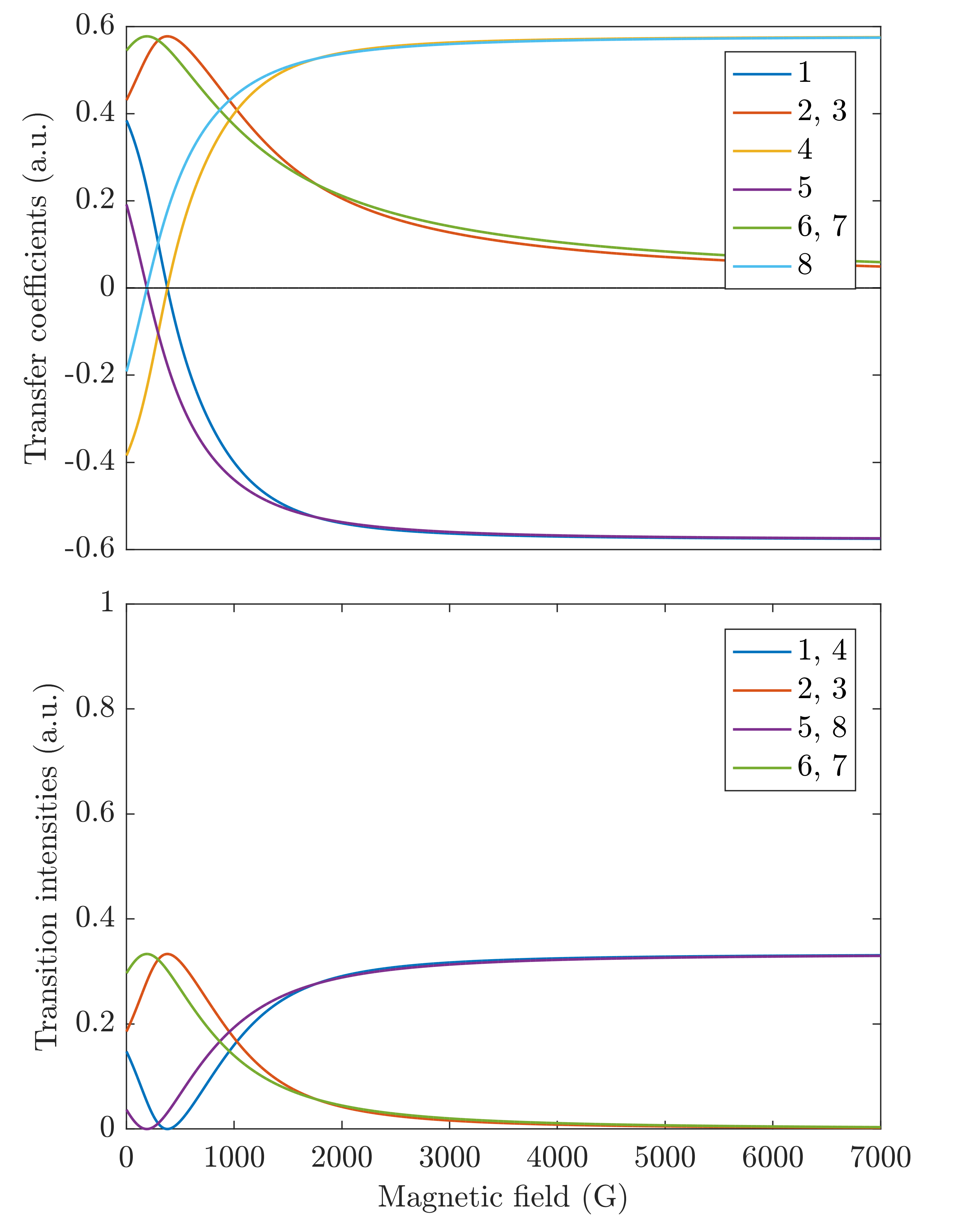}
\caption[]{$^{85}$Rb $D_1$ line $\pi$ transition transfer coefficients and transition intensities for $m=-2$ and $m=-1$.}
\label{fig:85_D1_pi}
\end{figure}

We have examined all the $\pi$, $\sigma^+$ and $\sigma^-$ transitions for $D_1$ line of $^{87}$Rb and $^{85}$Rb alkali atoms within magnetic field up to 10000~G. Indeed, as shown clearly on the Fig. \ref{fig:87_D1_pi}, \ref{fig:85_D1_pi}, \ref{fig:87_D2_pi}, \ref{fig:87_D2_sp}, \ref{fig:87_D2_sm}, all the graphs exhibit an asymptotic behavior after 5000~G or 6000~G. Thus no transition cancellation may be expect after these values, and all our graphs have been drawn up to the maximum value of 7000~G.

There are no $\sigma^+$ and $\sigma^-$ transition cancellations (i.e. transfer coefficients never become zero) for $^{87}$Rb and $^{85}$Rb $D_1$ line. From now on, we will give the results for $D_1$ line only for those groups of $\pi$ transitions (depending on $m$ value) which have at least one transition cancellation.

Our calculations were done with the values mentioned in Table~\ref{tab:D1_data}. One can notice, that the most imprecise value is the $\epsilon$ frequency difference between two excited state levels for both $^{87}$Rb and $^{85}$Rb isotopes. It has the biggest impact on the uncertainty size of the calculated magnetic field values.
\begin{table}[H]
\caption{Values used to calculate transfer coefficients and transition intensities of $\pi$ transitions for $^{87}$Rb and $^{85}$Rb $D_1$ line with their uncertainties.}
\begin{center}
\tabcolsep=0.15cm
\begin{tabular}{ccc}
\hline
Atom & Values & References \\
\hline
\multirow{6}{*}{$^{87}$Rb} & $\zeta=6 \; 834.682 \; 610 \; 904 \; 290(90)$ MHz & \cite{Bize_1999} \\\cline{2-3}
 & $\epsilon=814.50(13)$ MHz & \cite{Barwood1991,Banerjee,Arimondo} \\\cline{2-3}
 & $g_I=-0.000 \; 995 \; 1414(10)$ & \cite{Arimondo} \\\cline{2-3}
 & $g_S=2.002 \; 319 \; 304 \; 3622(15)$ & \cite{MohrCODATA2014} \\\cline{2-3}
 & $g_L=0.999 \; 993 \; 69$ & \cite{SteckRb87} \\\cline{2-3} 
 & $\mu_B/h=-1.399 \; 624 \; 5042(86)$ MHz/G & \cite{MohrCODATA2014} \\\hline
 \multirow{6}{*}{$^{85}$Rb} & $\zeta=3 \; 035.732 \; 439 \; 0(60)$ MHz & \cite{Arimondo,Budker2000} \\\cline{2-3}
 & $\epsilon=361.58(17)$ MHz & \cite{Barwood1991,Banerjee,Budker2000} \\\cline{2-3}
 & $g_I=-0.000 \; 293 \; 640 \; 00(60)$ & \cite{Arimondo} \\\cline{2-3}
 & $g_S=2.002 \; 319 \; 304 \; 3622(15)$ & \cite{MohrCODATA2014} \\\cline{2-3}
 & $g_L=0.999 \; 993 \; 54$ & \cite{SteckRb85} \\\cline{2-3} 
 & $\mu_B/h=-1.399 \; 624 \; 5042(86)$ MHz/G & \cite{MohrCODATA2014} \\\hline
\end{tabular}
\end{center}
\label{tab:D1_data}
\end{table}
On Fig.~\ref{fig:87_D1_pi} all possible $^{87}$Rb $D_1$ line $\pi$ transition transfer coefficients and transition intensities for $m=-1$ value are depicted.

Obviously all the graphs in the paper have been drawn without taking into account uncertainties of the involved quantities. Only two of four possible transitions have a cancellation and from \eqref{eq:B--} and \eqref{eq:B++} one can see that the analytical formulas for both transition cancellations are the same.
Taking into account the uncertainties of all the involved quantities in \eqref{eq:B--} and \eqref{eq:B++} allows us to determine the uncertainty of the magnetic field values. The analytically obtained value for the calculated above $^{87}$Rb $\pi$ transitions, for which the contribution of magnetic field cancels $\ket{F_g=1,m=-1}\rightarrow\ket{F_e=1,m=-1}$ and $\ket{F_g=2,m=-1}\rightarrow\ket{F_e=2,m=-1}$ transitions is $B=642.590(76)$ G. The values obtained by numerical simulation are shown in Table~\ref{tab:87_D1_pi}.
\begin{table}[H]
\caption{Magnetic field values cancelling $^{87}$Rb $D_1$ line $\pi$ transitions.}
\begin{center}
\begin{tabular}{cccccc}
\hline
No. & $F_{g}$ & $F_{e}$ & $m_{g}$ & $B$ (G) & $B^*$ (G)\\
\hline
1 & 1 & 1 & -1 & 642.590(76) & 642.5904743(48)\\\hline
4 & 2 & 2 & -1 & 642.590(76) & 642.5904743(48)\\\hline
\end{tabular}
\end{center}
\label{tab:87_D1_pi}
\end{table}
In Table~\ref{tab:87_D1_pi}, the numbers in the first column refer to the labeling of Fig.~\ref{fig:87_D1_pi}, second and third columns indicate ground and excited level $F$ total atomic angular momentum numbers respectively. Fourth column shows from which magnetic sublevel the transition occurs. Fifth column exhibits calculated magnetic field values taking into account all the uncertainties of the involved quantities. The $B^*$ value in the 6$^{\text{th}}$ column is obtained by ignoring the uncertainty on $\epsilon$. This calculation has been made in order to show how precise the $B$ values that cancel the transitions can be determined if this uncertainty on $\epsilon$ could be reduced. As immediate consequence, one sees the importance to determine experimentally an improved, i.e. more precise, value of $\epsilon$. From this very precisely known $\epsilon$ value, it becomes clear that $B^*$ could be considered as a new standard for magnetometer calibration.

It is very important to note that the analytically calculated values of magnetic field and the values which are obtained by numerical simulation are in very good agreement with each other. The adequacy of these two values is $10^{-12}$ and it means, that we will now use numerical simulation for $3 \times 3$ or $4 \times 4$ block matrices to find extremely precise values of magnetic field, which contributes to transition cancellations (instead of obtaining very complicated formulas that would need Cardano and Ferrari’s formulas).

\begin{figure}[H]
\begin{center}
\begin{tikzpicture}[scale=0.85]

\draw (0,5) -- (1,5);
\node [above] at (0.5,5) {$-3$};
\draw (1.2,5) -- (2.2,5);
\node [above] at (1.7,5) {$-2$};
\draw (2.4,5) -- (3.4,5);
\node [above] at (2.9,5) {$-1$};
\draw (3.6,5) -- (4.6,5);
\node [above] at (4.1,5) {$0$};
\draw (4.8,5) -- (5.8,5); 
\node [above] at (5.3,5) {$1$};
\draw (6,5) -- (7,5); 
\node [above] at (6.5,5) {$2$};
\draw (7.2,5) -- (8.2,5); 
\node [above] at (7.7,5) {$3$};
\node [right] at (8.4,5) {$F_e=3$};

\draw (1.2,3.8) -- (2.2,3.8) (2.4,3.8) -- (3.4,3.8) (3.6,3.8) -- (4.6,3.8) (4.8,3.8) -- (5.8,3.8) (6,3.8) -- (7,3.8);
\node [right] at (8.4,3.8) {$F_e=2$}; 

\draw (0,1.5) -- (1,1.5) (1.2,1.5) -- (2.2,1.5) (2.4,1.5) -- (3.4,1.5) (3.6,1.5) -- (4.6,1.5) (4.8,1.5) -- (5.8,1.5) (6,1.5) -- (7,1.5) (7.2,1.5) -- (8.2,1.5);
\node [right] at (8.4,1.5) {$F_g=3$}; 
\draw (1.2,0) -- (2.2,0) (2.4,0) -- (3.4,0) (3.6,0) -- (4.6,0) (4.8,0) -- (5.8,0) (6,0) -- (7,0);
\node [right] at (8.4,0) {$F_g=2$}; 

\draw [thick] [->] (1.4,0.1) -- (1.4,3.7);
\node [below] at (1.4,0) {1};
\draw [thick] [->] (1.6,0.1) -- (1.6,4.9);
\node [below] at (1.6,0) {2};
\draw [thick] [->] (1.8,1.6) -- (1.8,3.7);
\node [below] at (1.8,1.5) {3};
\draw [thick] [->] (2,1.6) -- (2,4.9);
\node [below] at (2,1.5) {4};

\draw [thick] [->] (2.6,0.1) -- (2.6,3.7);
\node [below] at (2.6,0) {5};
\draw [thick] [->] (2.8,0.1) -- (2.8,4.9);
\node [below] at (2.8,0) {6};
\draw [thick] [->] (3,1.6) -- (3,3.7);
\node [below] at (3,1.5) {7};
\draw [thick] [->] (3.2,1.6) -- (3.2,4.9);
\node [below] at (3.2,1.5) {8};

\draw [thick] [<->] (4.1,0.1) -- (4.1,1.4);
\node [right] at (4.1,0.75) {$\zeta'$};

\draw [thick] [<->] (4.1,3.9) -- (4.1,4.9);
\node [right] at (4.1,4.4) {$\epsilon'$};
\end{tikzpicture}
\caption[]{$^{85}$Rb $D_1$ line scheme in a magnetic field with $\pi$ transitions for $m=-2$ and $m=-1$.}
\label{fig:85_D1_pi_scheme}
\end{center}
\end{figure}
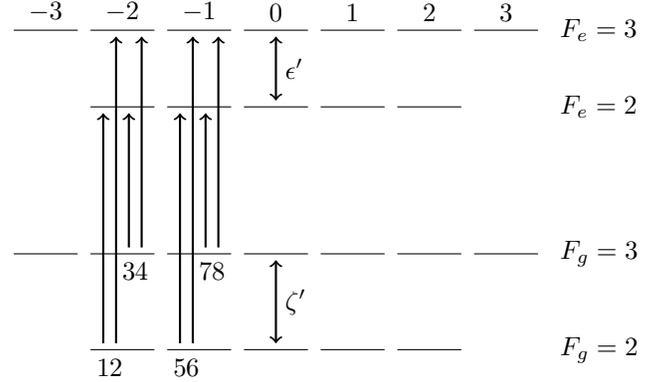

Next we examine $^{85}$Rb $D_1$ line $\pi$ transitions from $m=-2$ and $m=-1$ magnetic sublevels. Only these two groups of transitions contain some cancellations. It was mentioned before that no cancellation befalls for $\sigma^+$ and $\sigma^-$ transitions. The following scheme (Fig.~\ref{fig:85_D1_pi_scheme}) show grouped $\pi$ transitions depending on $m$ magnetic quantum number for this isotope.

Here we denote $\zeta'$ the frequency difference between the ground state $F_g=3$ and $F_g=2$ levels, and $\epsilon'$ the frequency difference between the $F_e=3$ and $F_e=2$ excited state levels. We obtained all analytical formulas for $^{85}$Rb $D_1$ line for magnetic field values, which cancel $\pi$ transitions:
\begin{equation}
B_{(\pm)}^{(\pm)}=\dfrac{1}{\mu_B} \cdot \dfrac{4\zeta'\epsilon'}{3 g_I \epsilon' -3 g_S \epsilon' + 3 g_I \zeta' -4 g_L \zeta' +g_S \zeta'}
\label{eq:B85-2}
\end{equation}
for $m=-2$ and
\begin{equation}
B_{(\pm)}^{(\pm)}=\dfrac{1}{\mu_B} \cdot \dfrac{2\zeta'\epsilon'}{3 g_I \epsilon' -3 g_S \epsilon' + 3 g_I \zeta' -4 g_L \zeta' +g_S \zeta'}
\label{eq:B85-1}
\end{equation}
for $m=-1$. Using these relations, we obtain $B=380.73(13)$~G, the value of the magnetic field which cancels transitions 1 and 4, as labelled on Fig.~\ref{fig:85_D1_pi_scheme}. Transitions 5 and 8 (see Fig.~\ref{fig:85_D1_pi_scheme}) are canceled for $B=190.368(66)$~G.
In Table~\ref{tab:85_D1_pi} are written all magnetic field values which cancel $^{85}$Rb $D_1$ line $\pi$ transitions. Again, in column 6, $B^*$ is calculated without taking into account the uncertainty on $\epsilon'$.
\begin{table}[H]
\caption{Magnetic field values cancelling $^{85}$Rb $D_1$ line $\pi$ transitions. First column indicates transfer coefficients and transition intensities according to the numeration shown on Fig.~\ref{fig:85_D1_pi}.}
\begin{center}
\begin{tabular}{cccccc}
\hline
No. & $F_{g}$ & $F_{e}$ & $m_{g}$ & $B$ (G) & $B^*$ (G)\\
\hline
1 & 1 & 1 &-2 & 380.73(13) & 380.7362466(29)\\\hline
4 & 2 & 2 & -2 & 380.73(13) & 380.7362466(29)\\\hline
5 & 1 & 1 & -1 & 190.368(66) & 190.3681233(15)\\\hline
8 & 2 & 2 & -1 & 190.368(66) & 190.3681233(15)\\\hline
\end{tabular}
\end{center}
\label{tab:85_D1_pi}
\end{table}

\section{Magnetic field values cancelling transitions of 87 and 85 Rubidium $D_2$ line}
\label{sec:D_2_line}
In this part we considered $D_2$ line transitions for both $^{87}$Rb and $^{85}$Rb isotopes. As mentioned before, we will present only those transitions, and respectively transfer coefficients, which have a cancellation. 

\subsection{$^{87}$Rb $D_2$ line}
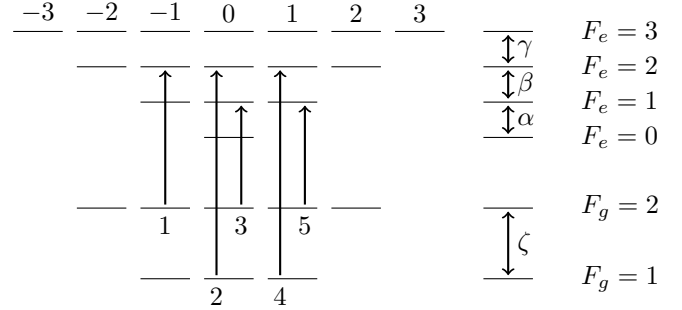
\begin{figure}[H]
\begin{center}
\begin{tikzpicture}[scale=0.47]

\draw (0,7) -- (1.4,7);
\node [above] at (0.7,7) {$-3$};
\draw (1.8,7) -- (3.2,7);
\node [above] at (2.5,7) {$-2$};
\draw (3.6,7) -- (5,7);
\node [above] at (4.3,7) {$-1$};
\draw (5.4,7) -- (6.8,7);
\node [above] at (6.1,7) {$0$};
\draw (7.2,7) -- (8.6,7); 
\node [above] at (7.9,7) {$1$};
\draw (9,7) -- (10.4,7); 
\node [above] at (9.7,7) {$2$};
\draw (10.8,7) -- (12.2,7); 
\node [above] at (11.5,7) {$3$};
\node [right] at (15.8,7) {$F_e=3$}; 

\draw (1.8,6) -- (3.2,6) (3.6,6) -- (5,6) (5.4,6) -- (6.8,6) (7.2,6) -- (8.6,6) (9,6) -- (10.4,6);
\node [right] at (15.8,6) {$F_e=2$}; 

\draw (3.6,5) -- (5,5) (5.4,5) -- (6.8,5) (7.2,5) -- (8.6,5);
\node [right] at (15.8,5) {$F_e=1$}; 

\draw (5.4,4) -- (6.8,4);
\node [right] at (15.8,4) {$F_e=0$}; 

\draw (1.8,2) -- (3.2,2) (3.6,2) -- (5,2) (5.4,2) -- (6.8,2) (7.2,2) -- (8.6,2) (9,2) -- (10.4,2);
\node [right] at (15.8,2) {$F_g=2$}; 
\draw (3.6,0) -- (5,0) (5.4,0) -- (6.8,0) (7.2,0) -- (8.6,0);
\node [right] at (15.8,0) {$F_g=1$}; 

\draw (13.3,0) -- (14.7,0) (13.3,2) -- (14.7,2) (13.3,4) -- (14.7,4) (13.3,5) -- (14.7,5) (13.3,6) -- (14.7,6) (13.3,7) -- (14.7,7);
\draw [thick] [<->] (14,0.1) -- (14,1.9);
\node [right] at (14,1) {$\zeta$};
\draw [thick] [<->] (14,4.1) -- (14,4.9);
\node [right] at (14,4.5) {$\alpha$};
\draw [thick] [<->] (14,5.1) -- (14,5.9);
\node [right] at (14,5.5) {$\beta$};
\draw [thick] [<->] (14,6.1) -- (14,6.9);
\node [right] at (14,6.5) {$\gamma$}; 

\draw [thick] [->] (4.3,2.1) -- (4.3,5.9);
\node [below] at (4.3,2) {1};
\draw [thick] [->] (5.75,0.1) -- (5.75,5.9);
\node [below] at (5.75,0) {2};
\draw [thick] [->] (6.45,2.1) -- (6.45,4.9);
\node [below] at (6.45,2) {3};
\draw [thick] [->] (7.55,0.1) -- (7.55,5.9);
\node [below] at (7.55,0) {4};
\draw [thick] [->] (8.25,2.1) -- (8.25,4.9);
\node [below] at (8.25,2) {5};
\end{tikzpicture}
\caption[]{$^{87}$Rb $D_2$ line scheme in magnetic field with all $\pi$ transitions which have a cancellation.}
\label{fig:87_D2_pi_scheme}
\end{center}
\end{figure}

We denote $\zeta$ the frequency difference between the ground state levels. For the excited state levels, notations are shown on Fig.~\ref{fig:87_D2_pi_scheme}. For $^{87}$Rb $D_2$ line only 5 $\pi$ transitions have a cancellation.

As already explained above, in this section we will not derive analytical formulas for the magnetic field values. For the numerical calculations we used values from Table~\ref{tab:D2_data}. Here too, all excited state levels frequency differences have relatively big uncertainties compared with others quantities involved in the calculations. In fact, this work can serve to determine more precisely excited state levels frequency differences. One of the possible techniques is to record selective reflection or/and transmission spectra. By making a fitting between theory and experiment it is possible to improve the following quantities: $\epsilon$ for $D_1$ and $\alpha$, $\beta$, $\gamma$ for $D_2$ line.
\begin{table}[H]
\caption{Excited state levels frequency differences for $^{87}$Rb and $^{85}$Rb $D_2$ line with their uncertainties.}
\begin{center}
\begin{tabular}{ccc}
\hline
Atom & Frequency difference (MHz) & References \\
\hline
\multirow{3}{*}{$^{87}$Rb} & $\alpha=72.218 0(40)$ & \multirow{3}{*}{\cite{Jun_Ye}} \\\cline{2-2}
 & $\beta=156.947 0(70)$ & \\\cline{2-2}
 & $\gamma=266.650 0(90)$ & \\\hline
 \multirow{3}{*}{$^{85}$Rb} & $\alpha'=29.372(90)$ & \multirow{3}{*}{\cite{Banerjee,Arimondo}} \\\cline{2-2}
 & $\beta'=63.401(61)$ & \\\cline{2-2}
 & $\gamma'=120.640(68)$ & \\\hline
\end{tabular}
\end{center}
\label{tab:D2_data}
\end{table}

In Table \ref{tab:D2_data} for $^{85}$Rb $D_2$ line $\alpha'$ is the frequency difference between $F_e=2$ and $F_e=1$ excited state levels, $\beta'$ is the frequency difference between $F_e=3$ and $F_e=2$ excited state levels and $\gamma'$ is the frequency difference between $F_e=4$ and $F_e=3$ excited state levels.

\begin{figure}[H]
\centering
\includegraphics[scale=0.7]{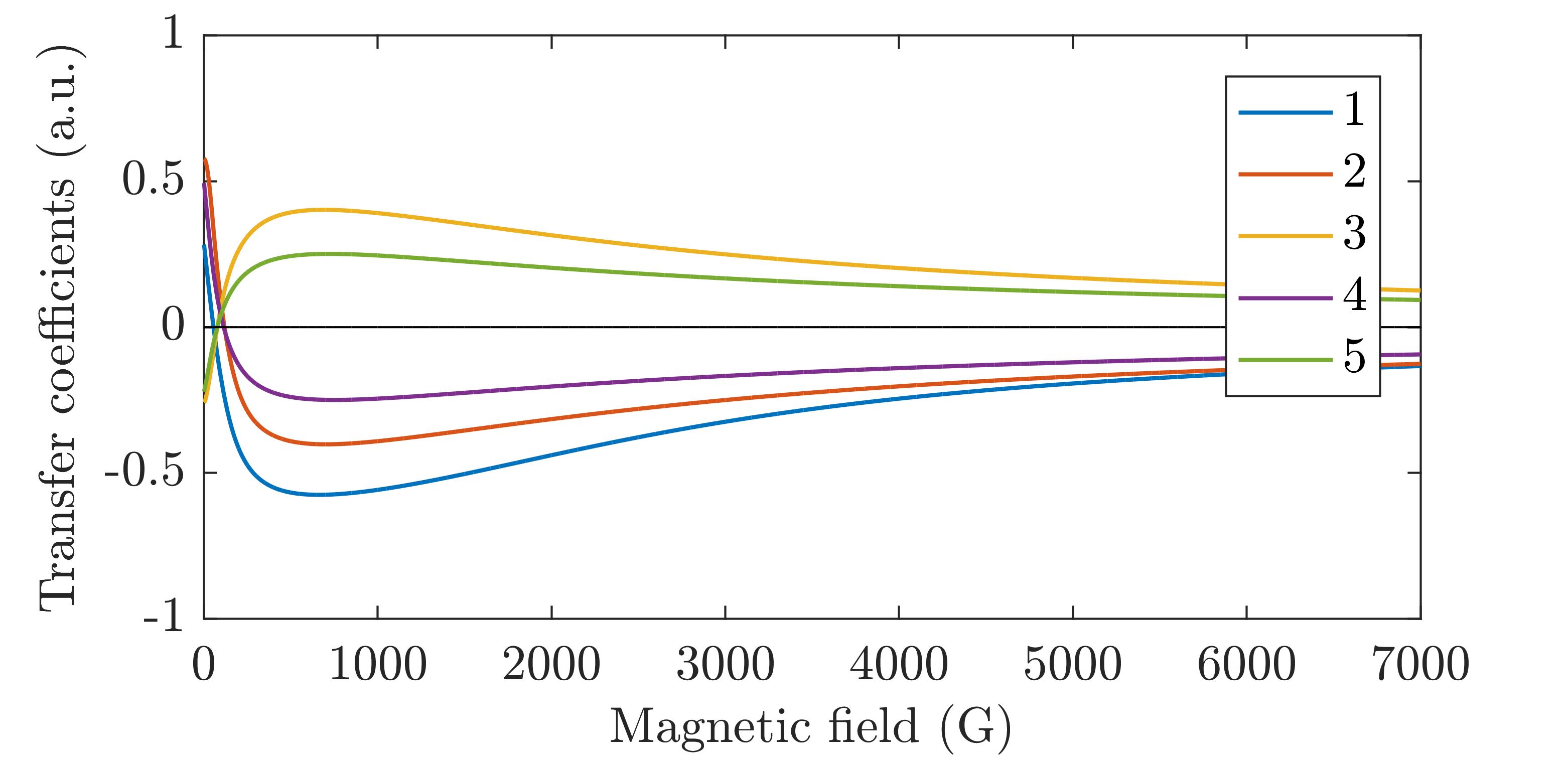}
\caption[]{$^{87}$Rb $D_2$ line $\pi$ transfer coefficients which have cancellation.}
\label{fig:87_D2_pi}
\end{figure}

\begin{figure}[H]
\centering
\includegraphics[scale=0.7]{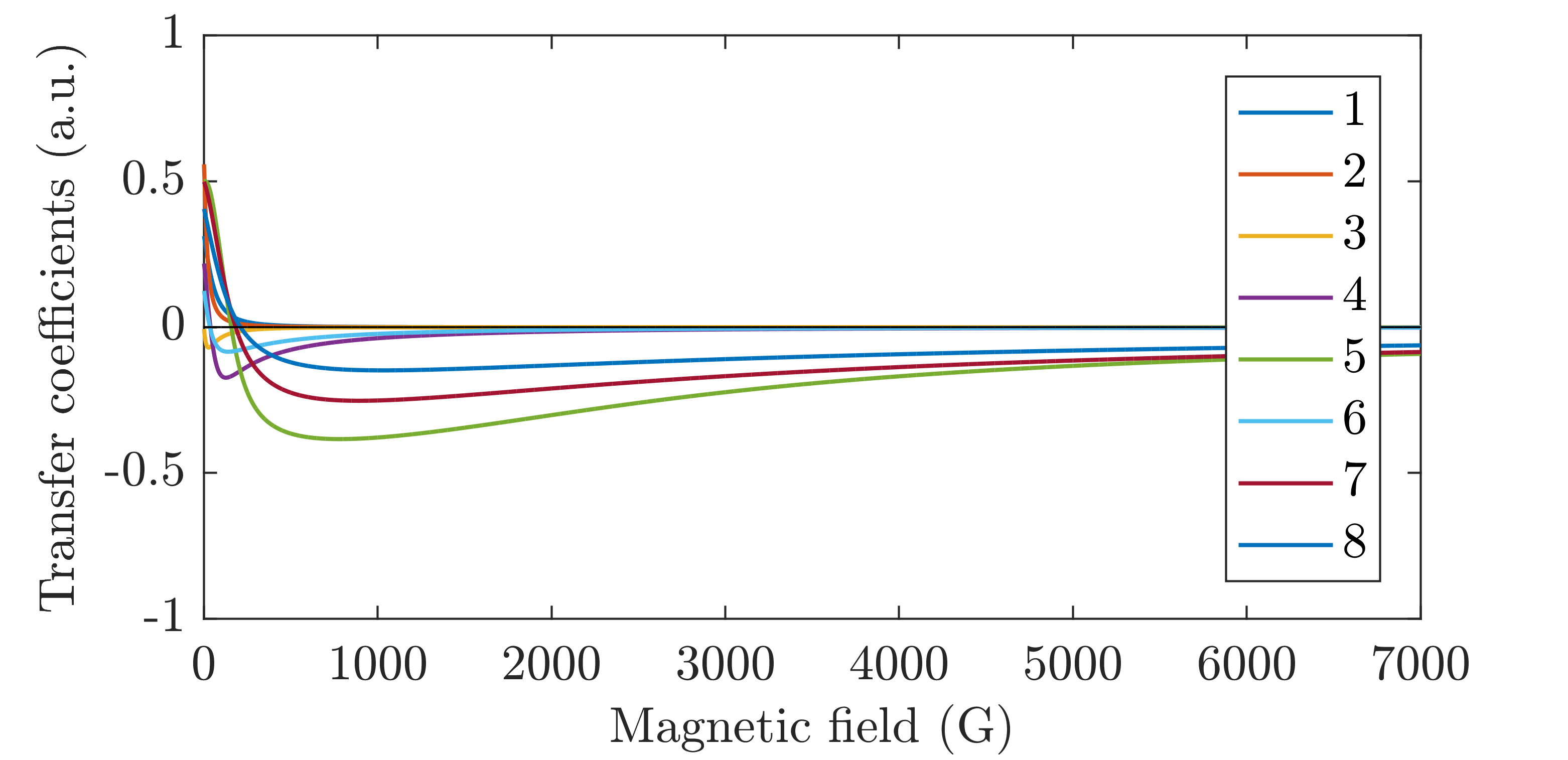}
\caption[]{$^{87}$Rb $D_2$ line $\sigma^+$ transfer coefficients which have cancellation.}
\label{fig:87_D2_sp}
\end{figure}

\begin{figure}[H]
\centering
\includegraphics[scale=0.7]{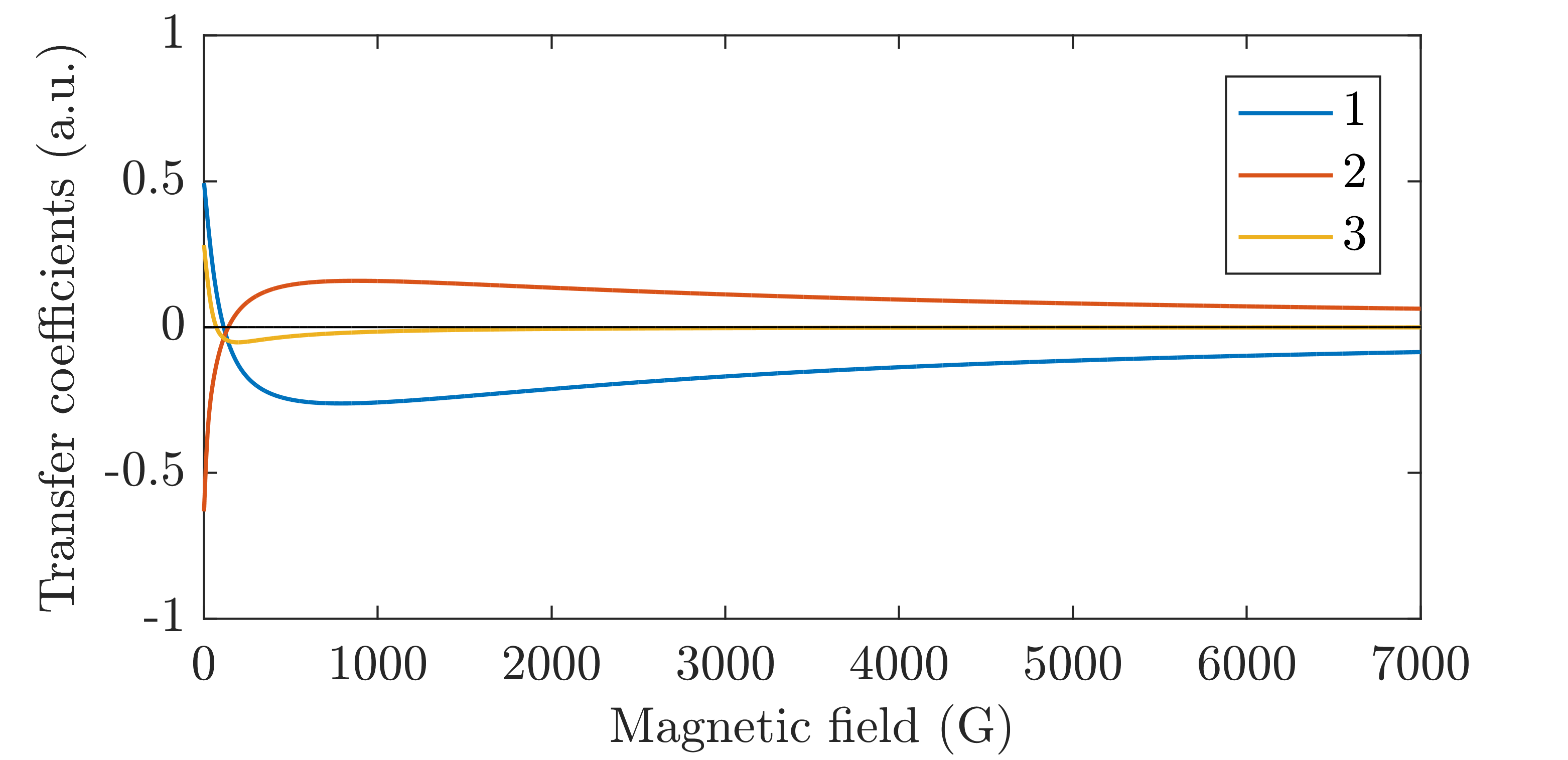}
\caption[]{$^{87}$Rb $D_2$ line $\sigma^-$ transfer coefficients which have cancellation.}
\label{fig:87_D2_sm}
\end{figure}

On Fig.~\ref{fig:87_D2_pi} are depicted all $\pi$ transitions for $^{87}$Rb $D_2$ line, which cancel for a certain value of magnetic field. For $D_2$ line there are no transitions which cancel for the same value of $B$, unlike the cases of $D_1$ line for both isotopes. This is visible on the figures \ref{fig:87_D2_pi}, \ref{fig:87_D2_sp} and \ref{fig:87_D2_sm}. One can see, that from 400~G, transfer coefficients become very small and for 7000~G the patterns of lines are very close to the asymptotic behavior.
In Table~\ref{tab:87_D2_pi} magnetic field values which cancel $\pi$ transitions are given. Again column 5 was calculated using all the uncertainties of involved quantities. Column 6 express magnetic field values without taking into account excited states uncertainties (i.e. we assume, that $\alpha$, $\beta$ and $\gamma$ have no uncertainties). Column 7 show on which frequency differences between excited state levels the uncertainty of magnetic field value depends on according to Fig.~\ref{fig:87_D2_pi_scheme}.
\begin{table}[H]
\begin{center}
\caption[]{Magnetic field values cancelling $\pi$ transitions for $^{87}$Rb $D_2$ line.}
\tabcolsep=0.15cm
\begin{tabular}{ccccccc}
\hline
No. & $F_{g}$ & $F_{e}$ & $m_{g}$ & $B$ (G) & $B^*$ (G) & $\Delta$E$_e$\\
\hline
1 & 2 & 2 & -1 & 55.6964(22) & 55.69646550(39) & $\beta$, $\gamma$\\\hline
2 & 1 & 2 & 0 & 118.7058(51) & 118.70586363(82) & $\alpha$, $\beta$, $\gamma$\\\hline
3 & 2 & 1 & 0 & 77.5048(35) & 77.50487199(54) & $\alpha$, $\beta$, $\gamma$\\\hline
4 & 1 & 2 & 1 & 114.2418(50) & 114.24183482(79) & $\beta$, $\gamma$\\\hline
5 & 2 & 1 & 1 & 77.2414(35) & 77.24147013(54) & $\beta$, $\gamma$\\\hline
\end{tabular}
\label{tab:87_D2_pi}
\end{center}
\end{table}
Below, on Fig.~\ref{fig:87_D2_sp_scheme} are shown all $^{87}$Rb $D_2$ line $\sigma^+$ transitions, which have a cancellation. There are only 8 transitions, and one of them, No.~3, is so-called forbidden. But due to the coupling of total atomic angular momenta ($F$) this transition become possible.
\begin{figure}[H]
\begin{center}
\begin{tikzpicture}[scale=0.47]
\draw (0,7) -- (1.8,7);
\node [above] at (0.9,7) {$-3$};
\draw (2.3,7) -- (4.1,7);
\node [above] at (3.2,7) {$-2$};
\draw (4.6,7) -- (6.4,7);
\node [above] at (5.5,7) {$-1$};
\draw (6.9,7) -- (8.7,7);
\node [above] at (7.8,7) {$0$};
\draw (9.2,7) -- (11,7); 
\node [above] at (10.1,7) {$1$};
\draw (11.5,7) -- (13.3,7); 
\node [above] at (12.4,7) {$2$};
\draw (13.8,7) -- (15.6,7); 
\node [above] at (14.7,7) {$3$};
\node [right] at (15.8,7) {$F_e=3$}; 

\draw (2.3,6) -- (4.1,6) (4.6,6) -- (6.4,6) (6.9,6) -- (8.7,6) (9.2,6) -- (11,6) (11.5,6) -- (13.3,6);
\node [right] at (15.8,6) {$F_e=2$}; 

\draw (4.6,5) -- (6.4,5) (6.9,5) -- (8.7,5) (9.2,5) -- (11,5);
\node [right] at (15.8,5) {$F_e=1$}; 

\draw (6.9,4) -- (8.7,4);
\node [right] at (15.8,4) {$F_e=0$}; 

\draw (2.3,2) -- (4.1,2) (4.6,2) -- (6.4,2) (6.9,2) -- (8.7,2) (9.2,2) -- (11,2) (11.5,2) -- (13.3,2);
\node [right] at (15.8,2) {$F_g=2$}; 
\draw (4.6,0) -- (6.4,0) (6.9,0) -- (8.7,0) (9.2,0) -- (11,0);
\node [right] at (15.8,0) {$F_g=1$}; 

\draw [thick] [->] (3.4,2.1) -- (5.3,4.9);
\node [below] at (3.4,2) {1};

\draw [thick] [->] (5,2.1) -- (7.6,5.9);
\node [below] at (5,2) {5};
\draw [thick] [->] (5.5,2.1) -- (7.4,4.9);
\node [below] at (5.5,2) {4};
\draw [thick] [->] (6,2.1) -- (7.2,3.9);
\node [below] at (6,2) {3};
\draw [thick] [->] (5.5,0.1) -- (8,3.9);
\node [below] at (5.5,0) {2};

\draw [thick] [->] (7.5,2.1) -- (10.15,5.9);
\node [below] at (7.5,2) {7};
\draw [thick] [->] (8.1,2.1) -- (10.05,4.9);
\node [below] at (8.1,2) {6};
\draw [thick] [->] (9.9,2.1) -- (12.4,5.9);
\node [below] at (9.9,2) {8};

\end{tikzpicture}
\caption[]{$^{87}$Rb $D_2$ line scheme in magnetic field with all $\sigma^+$ transitions which have a cancellation.}
\label{fig:87_D2_sp_scheme}
\end{center}
\end{figure}
Fig.~\ref{fig:87_D2_sp} demonstrates $^{87}$Rb $D_2$ line $\sigma^+$ transfer coefficients which cancel for a certain value of the magnetic field. Transfer coefficients are labeled accordingly with Fig.~\ref{fig:87_D2_sp_scheme}.
Table~\ref{tab:87_D2_sp} indicates magnetic field values which cancel certain $\sigma^+$ transitions. Column 5 ($B$) involves all the uncertainties of values used in the calculation. Column 6 ($B^*$) show magnetic field values without taking into account excited states uncertainties. In column 7 is written on which frequency differences between excited states the uncertainty of magnetic field value depends on according to Fig.~\ref{fig:87_D2_sp_scheme}. Magnetic field value, which cancel the $\ket{F_g=2,m=1}\rightarrow\ket{F_e=2,m=2}$ transition (No. 8), depends only on excited state $\gamma$ frequency difference between $F_e=3$ and $F_e=2$ levels.
\begin{table}[H]
\begin{center}
\caption[]{Magnetic field values cancelling $\sigma^+$ transitions for $^{87}$Rb $D_2$ line.}
\tabcolsep=0.15cm
\begin{tabular}{ccccccc}
\hline
No. & $F_{g}$ & $F_{e}$ & $m_{g}$ & $B$ (G) & $B^*$ (G) & $\Delta$E$_e$\\
\hline
1 & 2 & 1 & -2 & 1792.8(1.2) & 1792.854752(13) & $\beta$, $\gamma$\\\hline
2 & 1 & 0 & -1 & 1595.84(93) & 1595.846039(12) & $\alpha$, $\beta$, $\gamma$\\\hline
3 & 2 & 0 & -1 & 1762.3(1.7) & 1762.305097(13) & $\alpha$, $\beta$, $\gamma$\\\hline
4 & 2 & 1 & -1 & 37.7187(20) & 37.71876912(27) & $\alpha$, $\beta$, $\gamma$\\\hline
5 & 2 & 2 & -1 & 157.6244(63) & 157.6244550(11) & $\alpha$, $\beta$, $\gamma$\\\hline 
6 & 2 & 1 & 0 & 35.0323(19) & 35.03235682(25) & $\beta$, $\gamma$\\\hline
7 & 2 & 2 & 0 & 183.1469(71) & 183.1469403(13) & $\beta$, $\gamma$\\\hline
8 & 2 & 2 & 1 & 211.1182(80) & 211.1182479(15) & $\gamma$\\\hline
\end{tabular}
\label{tab:87_D2_sp}
\end{center}
\end{table}
Fig.~\ref{fig:87_D2_sm_scheme} shows the only cases of $\sigma^-$ $^{87}$Rb $D_2$ line transitions which cancel for a certain value of magnetic field. 
\begin{figure}[H]
\begin{center}
\begin{tikzpicture}[scale=0.47]
\draw (0,7) -- (1.8,7);
\node [above] at (0.9,7) {$-3$};
\draw (2.3,7) -- (4.1,7);
\node [above] at (3.2,7) {$-2$};
\draw (4.6,7) -- (6.4,7);
\node [above] at (5.5,7) {$-1$};
\draw (6.9,7) -- (8.7,7);
\node [above] at (7.8,7) {$0$};
\draw (9.2,7) -- (11,7); 
\node [above] at (10.1,7) {$1$};
\draw (11.5,7) -- (13.3,7); 
\node [above] at (12.4,7) {$2$};
\draw (13.8,7) -- (15.6,7); 
\node [above] at (14.7,7) {$3$};
\node [right] at (15.8,7) {$F_e=3$}; 

\draw (2.3,6) -- (4.1,6) (4.6,6) -- (6.4,6) (6.9,6) -- (8.7,6) (9.2,6) -- (11,6) (11.5,6) -- (13.3,6);
\node [right] at (15.8,6) {$F_e=2$}; 

\draw (4.6,5) -- (6.4,5) (6.9,5) -- (8.7,5) (9.2,5) -- (11,5);
\node [right] at (15.8,5) {$F_e=1$}; 

\draw (6.9,4) -- (8.7,4);
\node [right] at (15.8,4) {$F_e=0$}; 

\draw (2.3,2) -- (4.1,2) (4.6,2) -- (6.4,2) (6.9,2) -- (8.7,2) (9.2,2) -- (11,2) (11.5,2) -- (13.3,2);
\node [right] at (15.8,2) {$F_g=2$}; 
\draw (4.6,0) -- (6.4,0) (6.9,0) -- (8.7,0) (9.2,0) -- (11,0);
\node [right] at (15.8,0) {$F_g=1$}; 

\draw [thick] [->] (7.8,0.1) -- (5.5,5.9);
\node [below] at (7.8,0) {1};
\draw [thick] [->] (9.8,0.1) -- (7.7,4.9);
\node [below] at (9.8,0) {2};
\draw [thick] [->] (10.4,0.1) -- (7.9,5.9);
\node [below] at (10.4,0) {3};
\end{tikzpicture}
\caption[]{$^{87}$Rb $D_2$ line scheme in magnetic field with all $\sigma^-$ transitions which have a cancellation.}
\label{fig:87_D2_sm_scheme}
\end{center}
\end{figure}
$\sigma^-$ transition transfer coefficients, which have a cancellation for $^{87}$Rb $D_2$ line are depicted on Fig.~\ref{fig:87_D2_sm}. Lines on the figure are labeled in accordance with Fig.~\ref{fig:87_D2_sm_scheme}.
In Table~\ref{tab:87_D2_sm} magnetic field values for all possible $^{87}$Rb $D_2$ line $\sigma^-$ transition cancellations are expressed.
\begin{table}[H]
\begin{center}
\caption[]{Magnetic field values cancelling $\sigma^-$ transitions for $^{87}$Rb $D_2$ line.}
\tabcolsep=0.15cm
\begin{tabular}{ccccccc}
\hline
No. & $F_{g}$ & $F_{e}$ & $m_{g}$ & $B$ (G) & $B^*$ (G) & $\Delta$E$_e$\\
\hline
1 & 1 & 2 & 0 & 114.3072(50) & 114.30723113(80) & $\beta$, $\gamma$\\\hline
2 & 1 & 1 & 1 & 140.8256(71) & 140.82560775(98) & $\alpha$, $\beta$, $\gamma$\\\hline
3 & 1 & 2 & 1 & 71.9264(47) & 71.92641933(50) & $\alpha$, $\beta$, $\gamma$\\\hline
\end{tabular}
\label{tab:87_D2_sm}
\end{center}
\end{table}

\subsection{$^{85}$Rb $D_2$ line}
Hereafter we will examine $^{85}$Rb $D_2$ line $\pi$, $\sigma^+$ and $\sigma^-$ transfer coefficients within magnetic field. We will consider only transitions which have a cancellation. $^{85}$Rb $D_2$ line is a much more complicated system than $^{87}$Rb $D_2$ line, with large total atomic angular momentum ($F$) numbers. We will not show any scheme or transfer coefficients concerning transitions, because distinguishing one line from another would be very hard. We will bring only tables where magnetic field values which cancel certain transitions are indicated. As one can notice, for $^{85}$Rb $D_2$ line the frequency differences between excited state levels are smaller than in the case of $^{87}$Rb $D_2$ line. For some cases we obtained analytical formulas similar to \eqref{eq:B--}, \eqref{eq:B++}, \eqref{eq:B85-2} and \eqref{eq:B85-1}, where the value of $B$-field cancelling transitions mostly depends on excited and ground state level frequency differences (i.e. $\alpha'$, $\beta'$, $\gamma'$ and $\zeta'$). Because of that the values of $B$-field which cancel certain transitions are generally smaller than the $B$-field values obtained in the case of $^{87}$Rb $D_2$ line.

Table~\ref{tab:85_D2_pi} includes all magnetic field values which cancel certain $\pi$ transitions. Attentive readers may notice that some values of magnetic field that cancel certain transitions are too big. And accordingly, the uncertainties of these values are big too. There are no such results for $^{87}$Rb $D_2$ line. In order to improve uncertainties of the involved parameters, the first step is to try to measure more precisely the values of magnetic field which cancel these transitions. The second step includes in itself the measurement of magnetic field values for those transitions which uncertainties depends on only one frequency difference (e.g. last line of Table~\ref{tab:85_D2_sp}). So, by measuring different magnetic field values which cancel certain transitions, it is possible to decrease the uncertainties of excited state levels frequency differences.
\begin{table}[H]
\caption[]{$B$-field values cancelling $^{85}$Rb $D_2$ line $\pi$ transitions.}
\begin{center}
\begin{tabular}{cccccc}
\hline
$F_{g}$ & $F_{e}$ & $m_{g}$ & $B$ (G) & $B^*$ (G) & $\Delta$E$_e$\\
\hline
3 & 3 & -2 & 31.977(23) & 31.97774839(22) & $\beta'$, $\gamma'$\\\hline
2 & 2 & -1 & 6.565(17) & 6.565192522(44) & $\alpha'$, $\beta'$, $\gamma'$\\\hline
2 & 3 & -1 & 48.463(58) & 48.46368819(33) & $\alpha'$, $\beta'$, $\gamma'$\\\hline
2 & 4 & -1 & 5686(29) & 5686.364269(49) & $\alpha'$, $\beta'$, $\gamma'$\\\hline
3 & 2 & -1 & 35.228(43) & 35.22828802(24) & $\alpha'$, $\beta'$, $\gamma'$\\\hline
3 & 3 & -1 & 12.811(11) & 12.811030753(85) & $\alpha'$, $\beta'$, $\gamma'$\\\hline
2 & 3 & 0 & 47.491(54) & 47.49141288(32) & $\alpha'$, $\beta'$, $\gamma'$\\\hline
2 & 4 & 0 & 6013(29) & 6012.951766(52) & $\alpha'$, $\beta'$, $\gamma'$\\\hline
3 & 2 & 0 & 35.218(43) & 35.21852774(24) & $\alpha'$, $\beta'$, $\gamma'$\\\hline
2 & 3 & 1 & 46.336(49) & 46.33622671(31) & $\alpha'$, $\beta'$, $\gamma'$\\\hline
2 & 4 & 1 & 6345(29) & 6345.448972(54) & $\alpha'$, $\beta'$, $\gamma'$\\\hline
3 & 2 & 1 & 34.945(40) & 34.94502121(24) & $\alpha'$, $\beta'$, $\gamma'$\\\hline
2 & 3 & 2 & 45.099(42) & 45.09972813(31) & $\beta'$, $\gamma'$\\\hline
2 & 4 & 2 & 6681(30) & 6681.226747(57) & $\beta'$, $\gamma'$\\\hline
3 & 2 & 2 & 34.689(33) & 34.68962622(24) & $\beta'$, $\gamma'$\\\hline
\end{tabular}
\end{center}
\label{tab:85_D2_pi}
\end{table}
Table~\ref{tab:85_D2_sp} includes all magnetic field values up to 10000~G which cancel certain $\sigma^+$ transitions.
\begin{table}[H]
\caption[]{$B$-field values cancelling $^{85}$Rb $D_2$ line $\sigma^+$ transitions.}
\begin{center}
\begin{tabular}{cccccc}
\hline
$F_{g}$ & $F_{e}$ & $m_{g}$ & $B$ (G) & $B^*$ (G) & $\Delta$E$_e$\\
\hline
3 & 2 & -3 & 278.3(1.4) & 278.3151250(19) & $\beta'$, $\gamma'$\\\hline
2 & 1 & -2 & 180.9(1.5) & 180.9519212(13) & $\alpha'$, $\beta'$, $\gamma'$\\\hline
3 & 1 & -2 & 254.1(1.3) & 254.1070281(17) & $\alpha'$, $\beta'$, $\gamma'$\\\hline
3 & 2 & -2 & 16.798(26) & 16.79814373(12) & $\alpha'$, $\beta'$, $\gamma'$\\\hline
3 & 3 & -2 & 62.626(59) & 62.62663916(42) & $\alpha'$, $\beta'$, $\gamma'$\\\hline
2 & 1 & -1 & 156.9(1.6) & 156.9842182(11) & $\alpha'$, $\beta'$, $\gamma'$\\\hline
3 & 1 & -1 & 231.6(1.3) & 231.6749004(16) & $\alpha'$, $\beta'$, $\gamma'$\\\hline
3 & 2 & -1 & 15.983(23) & 15.98380527(11) & $\alpha'$, $\beta'$, $\gamma'$\\\hline
3 & 3 & -1 & 72.575(61) & 72.57573219(49) & $\alpha'$, $\beta'$, $\gamma'$\\\hline
2 & 1 & 0 & 137.2(1.6) & 137.21112478(91) & $\alpha'$, $\beta'$, $\gamma'$\\\hline
3 & 1 & 0 & 211.1(1.3) & 211.1105805(15) & $\alpha'$, $\beta'$, $\gamma'$\\\hline
3 & 2 & 0 & 15.337(20) & 15.33734519(11) & $\alpha'$, $\beta'$, $\gamma'$\\\hline
3 & 3 & 0 & 83.643(63) & 83.64378929(57) & $\alpha'$, $\beta'$, $\gamma'$\\\hline
3 & 2 & 1 & 14.808(18) & 14.80813301(10) & $\beta'$, $\gamma'$\\\hline
3 & 3 & 1 & 96.085(66) & 96.08519850(66) & $\beta'$, $\gamma'$\\\hline
3 & 3 & 2 & 110.162(71) & 110.16208826(76) & $\gamma'$\\\hline
\end{tabular}
\end{center}
\label{tab:85_D2_sp}
\end{table}
It is important to note, that the value written on last line of the table cancels $\ket{F_g=3,m=2}\rightarrow\ket{F_e=3,m=3}$ transition and its uncertainty depends only on excited state $F_e=3$ and $F_e=4$ levels frequency difference ($\gamma'$).

Table~\ref{tab:85_D2_sm} includes all magnetic field values for $^{85}$Rb $D_2$ line which cancel certain $\sigma^-$ transitions. One can notice that rows 4, 6 and 8 have two values for magnetic field cancelling one transition.
\begin{table}[H]
\caption[]{$B$-field values cancelling $^{85}$Rb $D_2$ line $\sigma^-$ transitions.}
\begin{center}
\begin{tabular}{cccccc}
\hline
$F_{g}$ & $F_{e}$ & $m_{g}$ & $B$ (G) & $B^*$ (G) & $\Delta$E$_e$\\
\hline
2 & 3 & -1 & 46.630(40) & 46.63046914(32) & $\beta'$, $\gamma'$\\\hline
2 & 4 & -1 & 4718(20) & 4718.168407(41) & $\beta'$, $\gamma'$\\\hline
2 & 2 & 0 & 50.440(68) & 44005212(34) & $\alpha'$, $\beta'$, $\gamma'$\\\hline
2 & 3 & 0 & \begin{tabular}{@{}c@{}} 32.361(41)\\ 4354(19)\end{tabular} & \begin{tabular}{@{}c@{}}
32.36112827(22)\\ 
4354.588882(38)\end{tabular} & $\alpha'$, $\beta'$, $\gamma'$\\\hline
2 & 2 & 1 & 51.930(93) & 51.93093445(35) & $\alpha'$, $\beta'$, $\gamma'$\\\hline
2 & 3 & 1 & \begin{tabular}{@{}c@{}}29.726(51)\\ 
4005(19)\end{tabular} & \begin{tabular}{@{}c@{}} 29.72652541(20)\\ 
4004.977769(35)\end{tabular} & $\alpha'$, $\beta'$, $\gamma'$\\\hline
2 & 2 & 2 & 52.27(12) & 52.27464320(36) & $\alpha'$, $\beta'$, $\gamma'$\\\hline
2 & 3 & 2 & \begin{tabular}{@{}c@{}} 27.764(58)\\ 
3669(21)\end{tabular} & \begin{tabular}{@{}c@{}}
27.76483242(19)\\ 
3669.632908(32)\end{tabular} & $\alpha'$, $\beta'$, $\gamma'$\\\hline
\end{tabular}
\end{center}
\label{tab:85_D2_sm}
\end{table}

\section{Experimental feasibility analysis}
\label{feasibility}
Calculations for the cancellation of transitions in a magnetic field in the framework of the proposed model were carried out based on physical constants and the values of the basic quantities characterizing the atomic system under consideration, available from the literature (see Tables \ref{tab:D1_data}, \ref{tab:D2_data}). In the case of a proper experimental implementation, an accurate measurement of the magnetic field corresponding to the cancelling of the optical transition will make it possible to determine exact values of the physical parameters, in particular the value of frequency difference between the upper state levels $\epsilon$, the only physical constant determined so far with least precision. Carefully elaborated experimental configuration and extremely high accuracy in measuring the applied magnetic field are required to achieve this goal, which makes the task ambitious. Let us briefly analyze the requirements to experimental setup and its characteristics needed for defining new physical constants standards.

First, in thermal atomic vapor the hyperfine transitions, and especially, transitions between the magnetic sublevels of hyperfine states are Doppler--broadened and overlapped. To work with a chosen individual transition, it has to be frequency--separated from the neighboring ones. This can be done with the use of high--resolution spectroscopic techniques providing sub--Doppler or Doppler--free frequency resolution, in particular, monokinetic atomic beam \cite{Peik, Pillet} or nanocell \cite{Papageorgiou1994,Hakhumyan2012} spectroscopy. Moreover, the tuning range of a single--frequency cw laser should be sufficiently large to follow the frequency shift of the chosen transition in a $B$-field. These requirements are easy to fulfill with the use of non--expensive diode lasers and Rb vapor nanocells with $\approx \lambda/2$ thickness in selective reflection configuration providing $\approx$~40~MHz linewidth \cite{Sargsyan2016}, or in the fluorescence configuration providing $\approx$~60~MHz linewidth \cite{Hakhumyan2010,Gazazyan2007}. These widths are sufficient for the complete separation of individual transitions, and hence the study of the cancellation, for magnetic fields above $\approx$~100~G. Noteworthy, both of these techniques assure a linear response of the atomic medium \cite{Sargsyan2016, Gazazyan2007}, unlike the widely used sub--Doppler technique of saturated absorption spectroscopy. The use of nanocells is advantageous also for a guaranteed uniformity of the applied magnetic field thanks to extremely small size of the interaction region \cite{Sargsyan2012, Sargsyan2017}.

Another important point is detection sensitivity. The precision of transition cancellation is physically limited by a noise level. Here the figure of merit is a signal--to--noise ratio (SNR). The level of typical selective reflection signal varies within $\approx$~5\% from the incident light signal. In contrast, the fluorescence signal has a zero off--resonance background. Conventional signal acquisition and processing techniques allow reliable detection of signal with SNR up to 10000. For particular cases of selective reflection and fluorescence measurements, the realistic estimate for the magnitude of cancelled transition is $\sim$~0.1\% of the initial ($B=0$) value.

Furthermore, the signal magnitude can be affected by the accuracy of setting and maintaining a given thickness of the nanocell in the interaction region. This problem is easily solved by controlling the radiation beam diameter and precise positioning of the beam with micro--controlled translation stage.

The main limitation are expected to come from the precision of application and measurement of a magnetic field. We should clearly distinguish two aspects: i) the accuracy of magnitude and direction of the applied $B$-field needed to cancel the transition, and ii) the precision of measurement of this field. We believe the most appropriate solution combining magnetic field control with its measurement may be the use of optical compensation magnetometry \cite{Papoyan2016}. The essence of the method is as follows. The interaction region, i.e. the vapor nanocell, is mounted into a system of calibrated Helmholtz coils (three mutually perpendicular pairs). Coil currents are scanned according to a special algorithm controlled by the studied transition signal. Using the method of successive approximations, a magnetic field value corresponding to the minimum of the atomic signal is achieved, and from the corresponding current values of coils currents a cancelling field value is determined. With the use of this method, control and measurement of a $B$-field with $\approx$~1~mG accuracy is experimentally feasible.

Last but not least, in the course of the measurements the laser radiation frequency should be stabilized on the transition under study. This can be done by implementing a feedback--based tunable locking of radiation frequency to an atomic resonance providing $\sim$~2~MHz accuracy \cite{Papoyan2004}, realized on an auxiliary setup with the second nanocell.

The above analysis shows that the expected realistic accuracy of the application and measurement of the magnetic field in the experiment is still far from the precision of the calculated values given in the tables of Sections \ref{sec:D_1_line} and \ref{sec:D_2_line}. However, it should be noted that, as indicated above, it is possible to decrease the uncertainties of excited state levels frequency differences by measuring the cancellation $B$-field values for different transitions, for which the uncertainties depend on one frequency difference (e.g. last line of Table \ref{tab:85_D2_sp}).

Besides a more accurate determination of physical quantities, the obtained results can be used for practical applications, in particular, for magnetometry and optical information. Continuous detection of an atomic signal while moving the nanocell across highly non--uniform magnetic field will allow a high--contrast optical mapping of a $B$-field. On the other hand, modulation of the magnetic field around the transition cancellation point will allow to modulate the amplitude of the optical atomic signal that carries optical information.

\section{Conclusion and outlook}
Summarizing, we have developed a precise model to calculate intensities of all the optical transitions between magnetic sublevels of hyperfine levels, excited with $\sigma^+$, $\pi$ and $\sigma^-$ polarized light, for the $D_1$ and $D_2$ lines of $^{87}$Rb and $^{85}$Rb atoms. Our analytical and numerical calculations have revealed complete cancelling of some individual transitions at certain, precisely determined values of the magnetic field that can serve as standardized quantities characterizing the atomic system.

We have calculated all the transition-cancelling $B$ values using two different methods. In the first method, all the parameters are kept with their uncertainties. The obtained magnetic field values are given in tables, and obviously the precision is strongly affected by the uncertainty of the excited state levels frequency differences. In the second method, the excited state levels frequency differences were used with their uncertainties supposed to be exact, while other parameters were supposed not to be exact (tables \ref{tab:87_D1_pi}, \ref{tab:85_D1_pi}, \ref{tab:87_D2_pi}, \ref{tab:87_D2_sp}, \ref{tab:87_D2_sm} column 6 and tables \ref{tab:85_D2_pi}, \ref{tab:85_D2_sp}, \ref{tab:85_D2_sm} column 5). These columns clearly indicate that the uncertainty on magnetic field value arises only from the excited state levels frequency differences.

We believe the appropriate experimental realization will allow reducing the uncertainties of some physical parameters, in particular the values of frequency difference between the upper state levels, that are currently determined with a least accuracy.

In addition, we have outlined other applications, notably in optical magnetometry and optical information, where the obtained results can be used.

$\newline$
\noindent\textbf{Funding.} Artur Aleksanyan acknowledges the funding support CO.17049.PAC.AN from the Graduate School EUR EIPHI.


    \bibliographystyle{IEEEtran}
       
    \bibliography{ArXiv_transition_cancellations_of_87_Rb_and_85_Rb_atoms_in_a_magnetic_field_setting_new_standards}


\end{multicols}
\end{document}